\documentclass[12pt]{article}
\usepackage{epsf}
\usepackage{latexsym}
\newcommand{\be}{\begin{equation}}
\newcommand{\ee}{\end{equation}}
\newcommand{\ba}{\begin{array}}
\newcommand{\ea}{\end{array}}
\newcommand{\bqa}{\begin{eqnarray}}
\newcommand{\eqa}{\end{eqnarray}}
\begin{document}
\begin{center}

{\Large\sf Left Hand Singularities,
           Hadron Form Factors and
           the Properties of the $\sigma$ Meson}
\\[10mm]
{\sc Zhiguang Xiao and Hanqing Zheng}
\\[5mm]
{\it Department of Physics, Peking University, Beijing 100871, P.~R.~China
}
\\[5mm]
\begin{abstract}
 By applying analyticity and single channel unitarity we
derive a new formula which is  useful to analyze the role of
the left--hand singularities  in hadron form factors and in the
determination of the  resonance parameters.  Chiral perturbation
theory is used to estimate the left--hand cut effects in $\pi\pi$
scattering processes. We find that in the $IJ=11$ channel the
left--hand cut effect is negligible and in the $IJ=20$ channel the
phase shift is dominated by the left--hand cut effect. In the
$IJ=00$ channel the left--hand cut contribution to the phase shift
has the wrong sign comparing with the experimental data and
therefore it necessitates the
$\sigma$ resonance. The new experimental results from
the E865 collaboration
is crucial in reducing the uncertainty in the determination
of the mass and width of the $\sigma$ resonance within our scheme.
\end{abstract}
\end{center}

\section{Introduction}
\label{SecIntr}

\begin{figure}
\mbox{\epsfxsize=12cm \epsffile{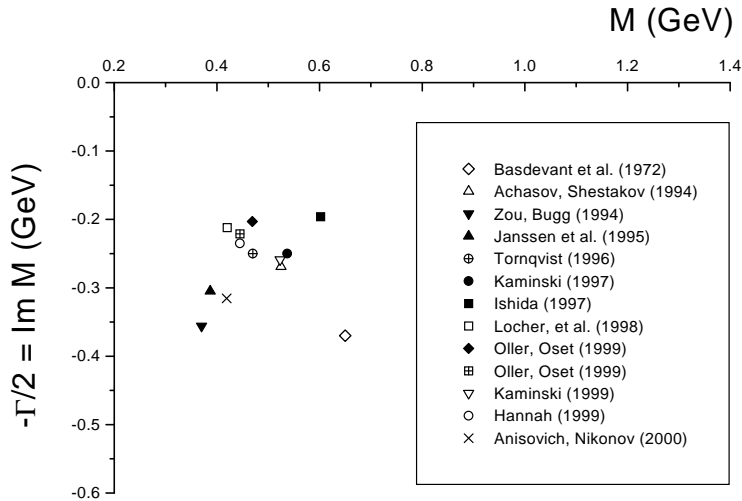}}
\caption{\label{Figsigma} The poles of the $S$-matrix in the
complex mass plane (GeV) corresponding to the lightest scalar
resonance according to Refs.~ \protect\cite{BFP72} --
\protect\cite{AN00}. }
\end{figure}

   The dynamical origin of the lightest scalar meson
is a subject of long lasting debates and controversies that were
reflected in the name $f_0(400-1200)$ given to it in latest PDG
publication \cite{RPP00,RPP98}.  There is a growing number of
studies claiming that the $\pi\pi$ scattering phase in the
scalar--isoscalar channel $J^{PC}=0^{++}$ $I=0$ which rises
steadily between the $\pi\pi$ threshold and the $f_0(980)$ region
is supported by a broad resonance known as the sigma meson. The
present situation is demonstrated in Fig.\ref{Figsigma} showing
the position of the $\sigma$ pole in the complex mass plane found
in different analysis
\cite{BFP72} -- \cite{AN00}. For most recent studies on related
issue, one is referred to Ref.~\cite{recentsigma}.

However, owing to the strong interaction nature, it is  difficult
to discuss the dynamics in the $IJ=00$ channel without heavily
relying on different models and approximations.
These models or approximations often violate some fundamental principles.
For example, in
most applications of the $K$ matrix approach the dynamical effects
of the left--hand cut ($l.h.c.$) are approximated by background
polynomials which are beyond control theoretically.
Strictly speaking, such an approximation violates
crossing symmetry.
It is not always clear how these assumptions and approximations
influence the conclusion at quantitative or even qualitative level.
Therefore it is very important to study the problem from different
angles and to reduce the model dependence as much as possible.

The goal of this paper is to elucidate the role of the left--hand
singularities in the determination of the position of the $\sigma$
meson. Our
philosophy in this paper is to start strictly from first
principles and to reduce model dependence as much as possible.
Throughout the text the main assumptions we use are analyticity
and single channel unitarity. We will also make use of chiral
perturbation theory to estimate the left--hand cut contributions
to $\pi\pi$ scattering processes in the phenomenological
application. Chiral perturbation theory encounters problems
at high energies,  an unitarized chiral approach in
studying the $\pi\pi$
system or even the $\pi\pi , \bar KK$ coupled--channel system has been
proposed and discussed
in recent years with remarkable success (see
Ref.~\cite{truong}--\cite{OOR00}, and also Ref.~\cite{Ha99}).
However, our treatment
of chiral perturbation theory is rather different from the standard
applications based on it. Essentially we only need
its prediction on cuts.
We are convinced, by analyzing the $IJ=00,20$ and
$11$ channels simultaneously, that chiral perturbation theory can
be used reliably to elucidate the role of the $l.h.c.$ in
the  determination of the $\sigma$ resonance.

This paper is organized as follows: In the begining we will
establish a new representation for a hadron form factor in the
single--channel case. We prefer to introduce it firstly in the
more traditional  scattering theory in sec.~\ref{SecFF}, which may
benefit some readers. The analytic structure of the $\pi$ hadronic
form factor and of the $\pi\pi$ scattering $S$ matrix in the
context of field theory is analyzed in the new scheme in
sec.~\ref{Sec-s-plane}. Left--hand cut  effects of the $\pi\pi$
scattering amplitudes are estimated in sec.~\ref{SecPionFF}. The
sec.~\ref{res} is devoted to study the pole position
of the $\sigma$ resonance. The analysis in the $IJ=20$ and $11$
channel is also made. The sec.~\ref{SecConcl} is for the
conclusion.

\section{The dispersion relation for a scalar form
factor}
\label{SecFF}

\subsection{Basic formulas}
\label{SecFFbasic}

In this section we consider the relationship between an $S$-matrix
and a scalar form factor in a single--channel scattering problem.
For the benefit of the reader, we begin with a brief summary
of well known properties of the scattering matrix.
The $S$-matrix is related to the partial wave amplitude
(we consider the case of the S-wave scattering) by
\bqa
   S(k) & = & 1 + 2i k f(k)
\label{Skf}
\eqa
where $k$ is the channel momentum.
The one--channel unitarity of the $S$ matrix in the physical region
$k \geq 0$ has the form
\bqa
   S(k)^* & = & S(k)^{-1} \quad ,
\label{Sunit}
\\
   \mbox{Im}\; f(k) & = &  k |f(k)|^2 \quad .
\label{funit}
\eqa
The {\it reflection property} of the $S$-matrix (see e.g. \cite{Newton})
leads to the following relation between positive and negative channel
momenta $k$:
\bqa
    S(-k) & = & S^*(k)
\label{Srefl}
\eqa
where the connection between $k<0$ and $k>0$ is via
an analytical continuation in the upper half plane.
The Riemann surfaces of the $S$-matrix and the scattering amplitude as
functions of energy $E=k^2$ have two sheets:
\bqa
    S(k) & = & \left\{ \ba{lcl}
    S_{I}(E) = 1+2i k_I(E) f_I(E)        & , & \mbox{Im}\;k \geq 0 \\
    S_{II}(E)= 1+2i k_{II}(E) f_{II}(E)  & , & \mbox{Im}\;k < 0
\ea
               \right.
\label{SE} \eqa Here $f(E)$ is the scattering amplitude, the
subscript $I$ or $II$ denotes its value on the sheet I or II, and
$k_I(E) = -k_{II}(E)$. In the following, however, we often drop
the subscript (or superscript) $I$ when it causes no confusion.
The reflection property (\ref{Srefl}) together with the unitarity
relation (\ref{Sunit}) gives the following relations between the
different branches of $S(E)$ and $f(E)$ \bqa
  S_{II}(E) & = & S_{I}^*(E) = \frac{1}{S_{I}(E)}
\\
\label{SISII}
    f_{II}(E) & = & \frac{f_I(E)}{S_{I}(E)}
\label{ffS}
\eqa
which extend by an analytical continuation from the physical
region $E\geq 0$ to the corresponding domains of analyticity on the
sheets I and II.

We define the scalar form factor $A(k)$ as an analytical function of $k$
which satisfies the following unitarity relation (compare with
Eq.(\ref{funit}))
\bqa
   \mbox{Im}\; A(k) & = &  k A(k) f(k)^*  \quad , \quad k\geq 0
\label{Aunit} \eqa and has no singularities, except for possible
bound states, in any finite part of the upper half plane
$\mbox{Im}\;k>0$ (that is the sheet I as the function of energy
$E$). To define the scalar form factor uniquely additional
constraints are needed. One is a trivial normalization condition
$A(k_0)=A_0$ with some choice of $k_0$ and $A_0$ convenient for
physical applications.

Equation (\ref{Aunit}) has the well known Omnes--Muskhelishvili
(OM) \cite{omnes} solution which determines the form factor
$A_I(E)$ on the sheet I \bqa
   A_I(E) & = & P(E) \exp{\left( \frac{E}{\pi}
                \int_{0}^{\infty} \frac{\delta(E')}{E'(E'-E)} dE'
\right)}
\label{OM}
\eqa
where $\delta(E)$ is the scattering phase:
\bqa
    S_I(E) & = &  e^{2i\delta(E)} \quad , \quad E\geq 0
\eqa and $P(E)$ is an arbitrary real polynomial ($P(E)$ is real
for real $E$). Here an once-subtraction form of the OM solution is
used, and the normalization condition is absorbed in the
polynomial $P(E)$.

\newcommand{\FJ}{\mbox{${\cal F}$}}
It  is useful to remember that in potential scattering the
$S$-matrix and the form factor can be conveniently defined in
terms of the Jost function $\FJ(k)$ \cite{Newton} \bqa
     S(k) & = & \frac{\FJ(-k)}{\FJ(k)} \label{SJF}
\\
     A(k) & = & \frac{1}{\FJ(k)}       \label{AJF}
\eqa
where $\FJ(k)$ is analytical in the upper half plane ($\mbox{Im}\;k>0$)
and has the reflection property:
\bqa
    \FJ(-k) & = & \FJ^*(k^*)
\eqa In this case the polynomial $P(E)$ is reduced to a constant.
Also one gets the relation between the values of the scalar form
factor on the sheets I and II: \bqa
    A_{II}(E) & = & \frac{A_I(E)}{S_I(E)}  \label{AIIAI}
\eqa

\subsection{A new representation of the scalar form factor}
\label{SecFFnew}

\begin{figure}
\mbox{\epsfxsize=10cm \epsffile{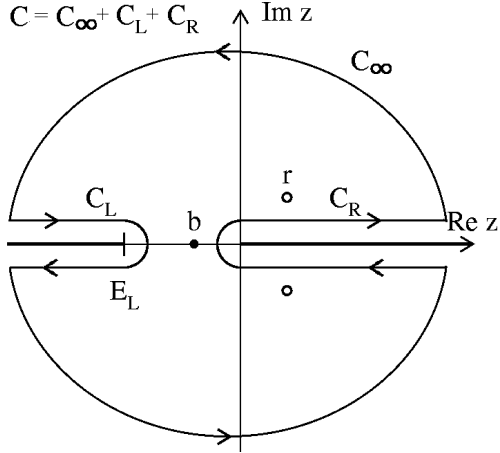}}
\caption{\label{FigContour}%
The integration contours used in the dispersion relations for
the scalar form factor.
}
\end{figure}

In this section we derive another useful representation of the
scalar form factor which, to the best of our knowledge, was
not described in the literature.
The starting point is the unitarity condition (\ref{Aunit}) which
is written in the form
\bqa
   \mbox{Im}\; A_I(E) = \frac{A_I(E+i\epsilon)-A_I(E-i\epsilon)}{2i}
               & = &  k_I(E) A_I(E) f_{II}(E)
\label{Aunit1} \eqa for $E \geq 0$. Here the identity
$f_{I}^*(E)=f_{II}(E)$ for real $E$ is used. Since the scalar form
factor $A_I(E)$ is analytical on the sheet I except for the poles
corresponding to bound states at $E<0$ and the kinematical cut
along the real $E$ axis, the following once--subtracted dispersion
relation can be used: \bqa
    A_I(E) - A_I(\Lambda)  & = &
             \sum_{b} \frac{(\Lambda-E) \beta_b}{(E-E_b)(\Lambda-E_b)} +
             I(E)\ ,
\label{ADR}
\\
    I(E)   & = &
           \frac{(E-\Lambda)}{2\pi i} \int_0^{\infty}
                             \frac{A_I(z+i\epsilon)-A_I(z-i\epsilon)}{(z-E)(z-\Lambda)} dz
\nonumber \\  & = &
           \frac{(E-\Lambda)}{2\pi} \int_{C_R}
                             \frac{k_I(z) A_I(z) f_{II}(z)}{(z-E)(z-\Lambda)} dz
\label{IE}
\eqa
where the sum is taken over all bound states with energies $E_b$,
$\beta_b$ are the corresponding residues of the form factor
in the poles on the Sheet I,
$\Lambda$ is a subtraction point,
and the contour $C_R$ envelopes the kinematical cut $0\leq E<\infty$ as shown
in Fig.\ref{FigContour}.

   The integration contour can be deformed, assuming that the
integrand falls off fast enough at the infinity (that is why the
subtraction is needed), in such a way that it goes around the {\it
left hand cut} $-\infty<E\leq E_c$ corresponding to the dynamical
singularities related to the scattering amplitude $f_{II}(E)$ on
the sheet II (see Fig.\ref{FigContour}). In the process of
deformation the contour crosses several poles of the integrand.
The terms $(z-E)$ and $(z-\Lambda)$ in the denominator of
Eq.(\ref{IE}) correspond to two such poles. The bound states, if
they exist, contribute via the poles on the sheet I in the factor
$A_I(E')$ . The virtual and resonant states correspond to the
poles on the sheet II of scattering amplitude $f_{II}(E')$.
Collecting all these terms we get
\bqa
I(E)  & = &  i (k_I(E) A_I(E) f_{II}(E) - k_I(\Lambda)
  A_I(\Lambda) f_{II}(\Lambda)
                 + B(E) + R(E) )
\\ & &
- \frac{(E-\Lambda)}{2\pi} \int_{C_L}
                    \frac{k_I(z) A_I(z) f_{II}(z)}{(z-E)(z-\Lambda)}
                    dz \ ,
\label{IELHC}
\\
   B(E) & = &
           \sum_{b} \frac{\beta_b k_I(E_b)f_{II}(E_b)(E-\Lambda)}{(E_b-E)(E_b-\Lambda)}\ ,
\label{BE}
\\
   R(E) & = &
           \sum_{r} \frac{ k(E_r) A_I(E_r)
           \mbox{Res}[f_{II}(E_r)](E-\Lambda)}{(E_r-E)(E_r-\Lambda)}\ .
\label{RE} \eqa In above the function $B(E)$ is given by the sum
over all bound states (the poles $E_b$ of the scattering amplitude
on the sheet I), and the function $R(E)$ is given by the sum over
all resonances and virtual states (the poles $E_r$ of the
scattering amplitude on the sheet II). According to
Eq.(\ref{SISII}) the poles of $S_I(E)$ correspond to the zeroes of
$S_{II}(E)$, therefore for all bound states $b$ \bqa
    k_I(E_b) f_{II}(E_b) & = & -\frac{i}{2}\ .
\label{fEj}
\eqa
The residues of $f_{II}(E)$ in the poles $E_i$ on the sheet II are
calculated with Eqs.(\ref{ffS},\ref{AIIAI}):
\bqa
    k_I(E_r) A_I(E_r) \mbox{Res}[f_{II}(E_r)] & = &
             \frac{i A_I(E_r)}{2 S'_{I}(E_r)} = \frac{i \beta_r}{2} \\
    \beta_r & = & \mbox{Res} A_{II}(E_r)
\label{ResfEi}
\eqa
Using Eqs.(\ref{ADR}-\ref{ResfEi}) we get
the following representation of the scalar form factor:
\bqa
     A_I(E) & = & \frac{S_I(E)}{1+S_I(E)} \left(
     A_I(\Lambda) \frac{1+S_I(\Lambda)}{S_I(\Lambda)}
  + \sum_{i=b,r} \frac{\beta_i}{E-E_i}- \sum_{i=b,r} \frac{\beta_i}{\Lambda-E_i}
\right. \nonumber \\ & & \left.
- \frac{(E-\Lambda)}{\pi} \int^{E_{L}}_{-\infty}
                   \frac{\Delta(E)}{(z-E)(z-\Lambda)} dz
               \right)
\label{AIE}
\eqa
where the function $\Delta(E)$ is defined on the left hand cut
$-\infty < E \leq E_L$ (see Fig.\ref{FigContour}) by
\bqa
     \Delta(E) & = &
     \frac{k_I(E) A_I(E) ( f_{II}(E+i\epsilon) - f_{II}(E-i\epsilon) )}{2i} \\
     & = &
     \frac{A_{II}(E+i\epsilon) - A_{II}(E-i\epsilon)}{2i} \label{DAII}
\eqa
Here we used Eq.(\ref{AIIAI}) in deriving the identity (\ref{DAII}).

The Eq.(\ref{AIE}), though very simple, has the very attractive
feature that it explicitly shows the contributions resulting from
different types of dynamical singularities: the bound states, the
resonances, the virtual states, and the left hand cut.  The poles
of the scattering amplitude contribute only to the {\it local}
terms in the representation (\ref{AIE}). The sum on the $r.h.s.$
of Eq.~(\ref{AIE}) includes the poles on both the first and the
second sheets. However, the poles on the sheet II are canceled by
the corresponding zeros of the S-matrix on the sheet I, so that
the representation (\ref{AIE}), which determines the scalar form
factor on the sheet I,  contains only the poles corresponding to
the bound states.

\subsection{Examples}
\label{SecFFex}

\begin{figure}
\begin{center}
\mbox{(a) \hspace{8cm} (b)}\\
\mbox{
\mbox{\epsfxsize=8cm \epsffile{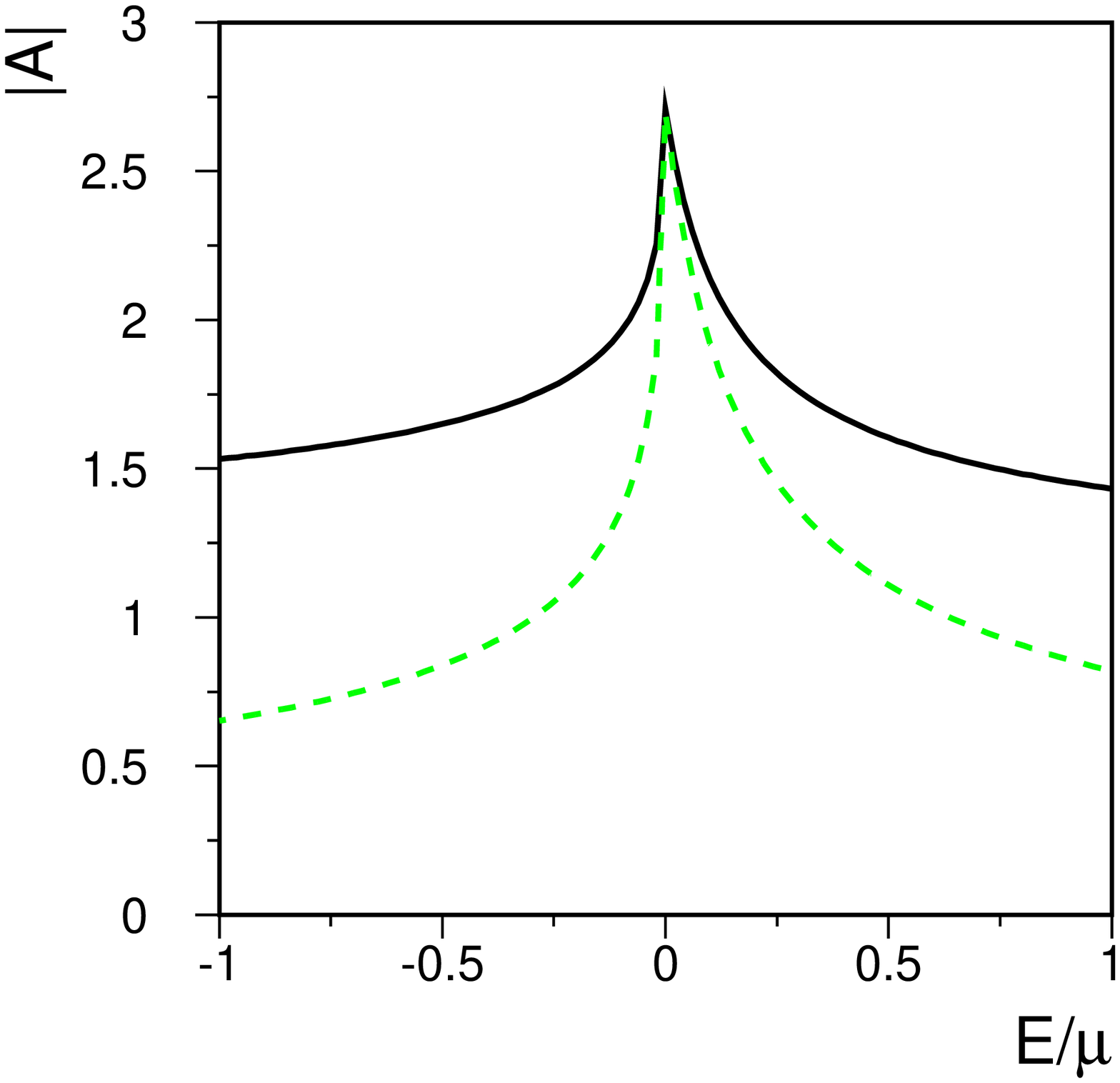}}
\mbox{\epsfxsize=8cm \epsffile{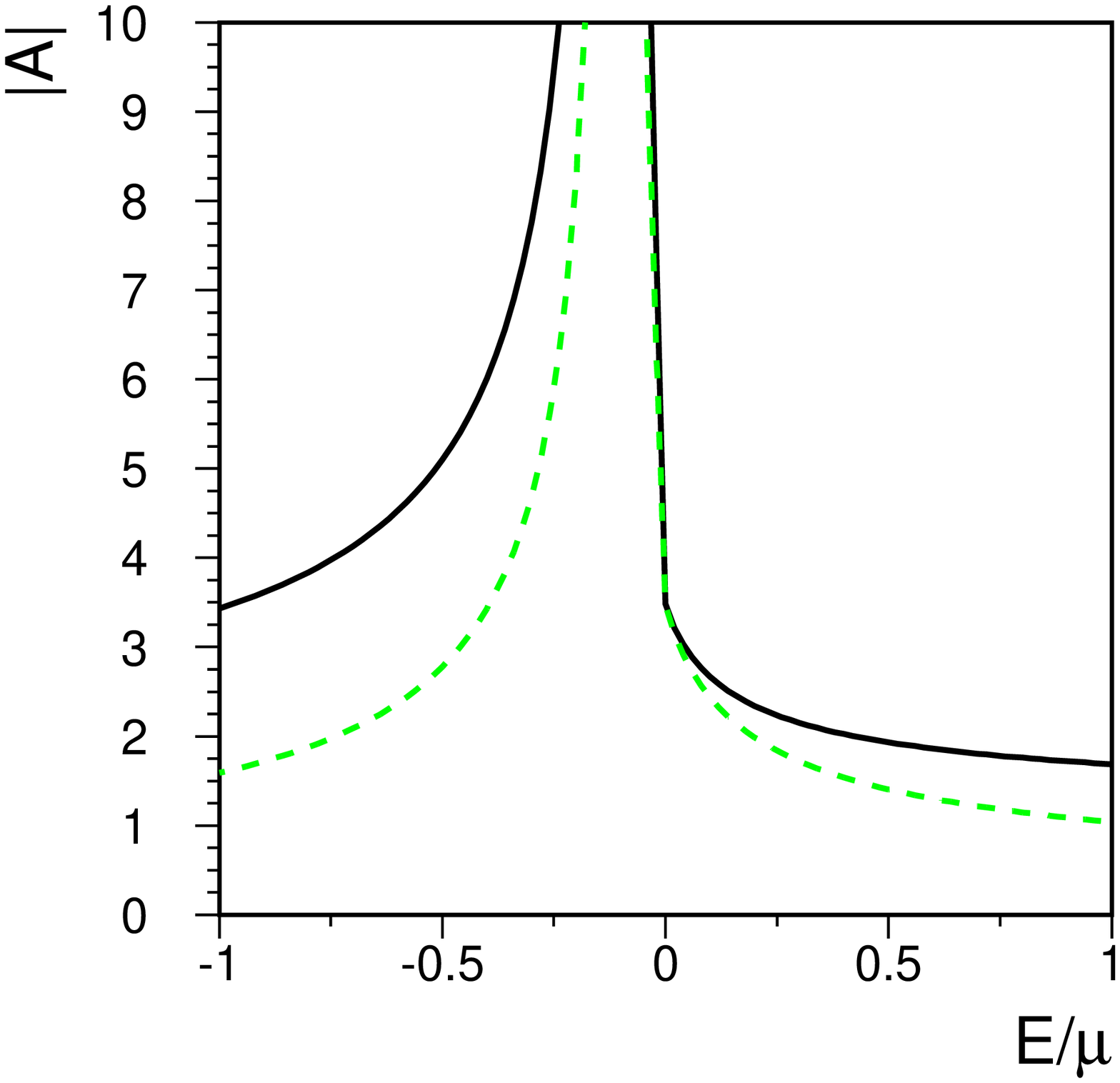}}
}
\end{center}
\caption{\label{FigYukawa}%
The scalar form factor corresponding to the Jost function defined
by Eq.(\protect\ref{FY1}): (a) a virtual state at $k=-i0.1\mu$,
(b) a bound state at $k=i0.1\mu$. The solid line is the complete
result for $|A_I(E)|$, the dashed lines show the result in the
scattering length approximation. The particle mass is set to
$m=0.5$, so that $E=k^2$. }
\end{figure}

To illustrate the above described formalism we consider a simple
example of a non-relativistic scattering of a particle with mass
$m$ by the Yukawa potential $V(r) = g \exp{(-\mu r)}/r$. The Jost
function in the first order in the coupling constant has the form
\bqa
     \FJ{(k)} & = & 1 - \frac{gm}{ik} \ln{(1-\frac{2ik}{\mu})}
\label{FY1} \eqa For the sake of simplicity, we neglect the higher
order terms because all analytical properties of the scattering
matrix and the form factor are strictly preserved in any finite
order
of the Jost function expansion.
\footnote{%
The expansion of $\FJ{(k)}$ in powers of the coupling constant is
straightforward.
} The S--matrix corresponding to the Jost function (\ref{FY1}) for
$g<0$ has only one pole at $k=i\kappa$: it is either a virtual
state on the sheet II ($\kappa<0$) or a bound state on the sheet I
($\kappa>0$). The calculations using Eq.(\ref{AIE}) are
straightforward, and typical examples for the cases of virtual and
bound states are shown in Fig.~\ref{FigYukawa}.

   One commonly used approximation is to neglect explicit
effects of the left hand cut and consider only a few poles of
the scattering amplitude close to the physical region.
In the scattering length approximation the S-matrix has the form
\bqa
     S_{a}(k) & = & \frac{1-ika}{1+ika}
\eqa
where $a$ is the scattering length.  In this case we have only the pole term
in (\ref{AIE}) at $2mE=k^2 = a^{-2}$, and the subtraction term vanishes for
$\Lambda\to\infty$ ($S(\infty)=-1$).  This gives the well known result:
\bqa
    A_{a}(E) & = & \frac{A(0)}{1+ik(E)a}
\label{Ascl}
\eqa
A comparison of this result with the above
considered example using $a = \pm \kappa^{-1}$ is shown in
Fig.~\ref{FigYukawa}. The range of validity of the scattering
length approximation for a weakly bound state is limited by the
energy range $|E|\sim 2m/a^2$. Note, that the approximation
Eq.~(\ref{Ascl}) does not have the correct asymptotic behaviour:
$A(E) \stackrel{E\to\infty}{\to}1$.

Another useful example is the pole approximation for $K$ matrix.
In the single--pole approximation, the $K$ matrix has the form
\bqa
     K(E) & = & \frac{\gamma_1}{E_1-E}
\label{KE}
\eqa
and the corresponding $S$ matrix is
\bqa
     S(E) & = & \frac{1+ik(E)K(E)}{1-ik(E)K(E)} \quad .
\label{SKE}
\eqa
Assuming $E_1>0$ and $0<\gamma_1<2\sqrt{E_1}$, one finds two resonant poles at
$E = E_r$ and $E=E_r^*$ where
$E_r = (-i\gamma_2 /2 + \sqrt{E_1-\gamma_1^2/4})^2$.
Using Eq.~(\ref{AIE}) we get
\bqa
     A_I(E) & = & \frac{S_I(E)}{1+S_I(E)}
                  \left( \frac{a}{E-E_r} + \frac{a^*}{E-E_r^*} \right ) =
\\
     & = &  \frac{(E_1-E+ik_I(E)\gamma_1)}{2(E_1-E)} \frac{C(E-E_z)}
{(E-E_r)(E-E_r^*)}
\eqa
where $a$ is the residue of the form factor at $E=E_r$ and $C$ is a normalization
constant.  At $\gamma_1 \neq 0$, the form factor has no pole
at the bare pole position
$E=E_1$, therefore $E_z=E_1$, and the form factor is given by
\bqa
     A_I(E) & = & \frac{(E_1-E+ik_I(E)\gamma_1)E_1}{(E-E_r)(E-E_r^*)}
\label{AIKE} \eqa where we set $C=2E_1$. The $r.h.s.$ of
(\ref{AIKE}) has no poles on the {\it first} energy sheet because
the nominator $(E_1-E+ik_I(E)\gamma_1)=(E_1-E-ik_{II}(E)\gamma_1)$
vanishes at $E = E_r, E_r^*$.  It is easy to verify that the
result (\ref{AIKE}) is equivalent the standard one pole
approximation~\cite{barton}: \bqa
     A_I(E) & = & \frac{E_1}{(E_1-E-ik_I(E)\gamma_1)}
\label{AISTE}
\eqa
This result can be easily generalized to the case when $K$ matrix has more than
one pole.

\section{The analytic form factor and the $S$ matrix in the complex $s$ plane}
\label{Sec-s-plane}

In the previous section we have introduced the new representation
for the form factor in the complex k--plane in the language of
scattering theory. It is of course straightforward to re-express
all the results in the above section in  relativistic quantum
field theory, in the complex $s$ plane (here $s$ denotes the
center of mass energy squared). Hence the spectral representation
of the form factor can be written down using the well known LSZ
reduction formalism,
\begin{equation}\label{sr}
{\rm Im}A=A\rho T^{*}\ ,
\end{equation}
 where $\rho$ denotes the kinematic phase-space factor. Taking
$\pi\pi$ scattering for example,
\begin{equation}
\rho=\sqrt{1-4m_\pi^2/s}\ .
\end{equation}
In Eq.~(\ref{sr})
$T$ denotes the  scattering T matrix which satisfies the
optical theorem,
\begin{equation}\label{sr2}
{\rm Im}T=T\rho T^{*}\ ,
\end{equation}
for the physical  value of $s$.  Discussions parallel to that in
section~\ref{SecFFnew} can be made and Eq.~(\ref{AIE}) can be
recasted as, \bqa\label{fa}
A_I(s)&=&{\frac{S_I}{1+S_I}}\{\sum_i{\frac{A_I(z_i)}{(s-z_i)
S_I^{\prime }(z_i)}} +\sum_j{\frac{\beta _j}{s-s_j}}+P_n(s)
\nonumber \\&& +{\frac{1}{2\pi i}}\int_L {\frac{A_I
(s^{\prime})disc\left [{\frac{1+S_I(s^{\prime})}{S_I(s^{\prime})
 }}\right ]}{s^{\prime}-s}}ds^{\prime}\ \}\ ,
\label{ff}
\eqa
where the position of zeros of $S_I$ on the complex $s$ plane
are denoted
as $z_i$, and $s_j$ are the position of bound state poles of
$S_I$ (and $A$).
The $P_n(s)$ is a (n-1)th order real polynomial when $A(s)$ obeys
a n-th subtracted dispersion relation. The discontinuity of $S$ on
the $l.h.c.$ $L=(-\infty, 0]$ manifests itself in the left--hand
integral on the $r.h.s.$ of the above equation, even though $A$
itself does not contain left--hand singularities. Possible (n-th)
subtractions on the integral in Eq.(\ref{ff}) are understood.

The Eq.~(\ref{ff}) offers a convenient and powerful expression for
studying the analytic structure of the scattering $S$ matrix and
form factors.
For example, it sets up a relation between
the form factor in the time-like region and in the space-like region,
provided that the $S$ matrix is known.
However, we will leave it aside
(except for the  analysis made under some simple approximations
discussed in sec.~\ref{SecFFex}) and go directly
to discuss the analytic property of the scattering T matrix.
In fact, it is easy to understand that the T matrix
itself satisfies Eq.~(\ref{ff}) with only minor modifications.
That is, unlike the form factor,
 $T$  also contains left--hand singularities.
 After some algebraic manipulation
one can obtain (for the detail of the proof, see Appendix) ,
\bqa
\label{FL} T(z)={S\over 1+S} \{ \sum_i {\frac{i
}{2\rho(z_i)S^{\prime}(z_i)(z-z_i)}} + \sum_j {\frac{ \beta_j}{z-s_j}} +
P_n(z)\nonumber \\ +{\frac{1}{2\pi i}}\int_L
{\frac{{disc}\left[T(s^{\prime}){\frac{
1+S(s^{\prime})}{S(s^{\prime})}}\right]
}{s^{\prime}-z}}ds^{\prime}\}\ .
\eqa
The relation
$T_I(z_i)=i/(2\rho(z_i))$ is used in deriving Eq.~(\ref{FL}). If we
define $F\equiv T(1+S)/S$ so the above equation can be rewritten
as,
\begin{equation}  \label{FL2}
F(z)= \sum_i {\frac{i/2\rho(z_i) }{S^{\prime}(z_i)(z-z_i)}} +
\sum_j {\frac{ \beta_j}{z-s_j}} + \\ P_n(z)+{\frac{1}{\pi }}\int_L
{\frac{{\rm Im} F }{s^{\prime}-z}}ds^{\prime}\ \ .
\end{equation}
The Eq.~(\ref{FL2}) sets up a dispersion relation for $F$ which
contains no right--hand cut.
 We can
further express the $S$ matrix in terms of F,
\begin{equation}  \label{S}
S=\sqrt{1-\rho^2 F^2}+i\rho F\ .
\end{equation}
Apparently $F$ is real when $z$ real and greater than $4m_\pi^2$.
Actually ${\rm Re} T={\frac{F}{2}}$ in the physical region, hence
$F$ is the analytic continuation of 2${\rm Re} T$ on the entire
cut plane, and now it becomes obvious why in Eq.~(\ref{FL2}) there
is no right--hand cut. Also we have,
\begin{equation}\label{IM}
{\rm Im}_R T={\frac{1-\sqrt{1-(\rho F)^2}}{2\rho}}\ ,
\end{equation}
for  physical value of $s$. It is worth pointing out that
Eq.~(\ref{IM}) relates the imaginary part of a given partial--wave
amplitude on the unitarity cut to the imaginary part of the same
partial--wave amplitude on the $l.h.c.$, in a complicated manner.
It is interesting to compare Eq.~(\ref{IM}) with the following
relation obtained using the Froissart--Gribov representation for
the partial wave projection,
\begin{eqnarray}
{\rm Im}_LT^I_l(s)&=&{\left[1+(-1)^{I+L}\right]\over
s-4m_\pi^2}\sum_{I'}\sum_{l'}(2l'+1)C_{II'}^{(st)}\nonumber \\
&&\times \int_{4m_\pi^2}^{4m_\pi^2-s}dtP_l(1+{2t\over
s-4m_\pi^2})P_{l'}(1+{2s\over t-4m_\pi^2}){\rm Im}T^{I'}_{l'}(t)
\end{eqnarray}%
which~\cite{MMS1} relates the left--hand discontinuity to the
right--hand discontinuity in a linear way, but all partial--waves
are involved. To further analyze Eq.~(\ref{S}), we notice that
$|\rho F(s)|\leq 1$ in the physical region, since in the physical
region
\be\label{delta} \rho F(s)=\sin(2\delta)\ .
\ee
It is therefore easy to understand that at  values of $s$ when
$\delta=\pi/4+n\pi/2$ we have,
\begin{equation}  \label{SR}
{\frac{d}{ds}}(\rho F) = 0 \ ,
\end{equation}
and
\begin{equation}  \label{ieq}
{\frac{d^2}{ds^2}}(\rho F) \le 0 \,\,\, (\rho F=1)\ ; \,\,\, {\frac{d^2}{ds^2%
}}(\rho F) \ge 0 \,\,\, (\rho F=-1)\ .
\end{equation}
These requirements lead to additional constraint on the parameters
in the expression of the $S$ matrix. Especially Eq.~(\ref{SR}) and
Eq.~(\ref{ieq}) indicate that the pole parameters and the
discontinuity on the left are correlated to each other. For
example, without further knowledge on the $l.h.c.$ integrals,
Eq.~(\ref{SR}) (Eq.~(\ref{ieq})) enables us to establish a sum
rule (inequalities) that the left--hand integral has to obey.

The Eq.~(\ref{delta}) is important in the sense that it relates
the experimental observable (the partial--wave phase shift) to the
pole contributions and the $l.h.c.$ contributions in a simple and
elegant manner. Unlike the conventional $K$ matrix approach, the
pole parameters are physical and the two contributions from poles
and $l.h.c.$ are additive. As we will see in the next section that
we will be able to calculate the $l.h.c.$ integral in
Eq.~(\ref{FL2}) in a reliable way, hence  Eq.~(\ref{delta}) offers
a convenient method to fit the pole positions. It is worth
emphasizing that Eqs.~(\ref{FL2}) and (\ref{delta}) are valid for
any partial--wave amplitude hence they afford a unified approach
for determining resonances in different channels. This is
remarkable since the
 output from different channels can be compared to each other and
this feature is very important to evaluate the quality of the
method and approximations being used in the fit.

\section{Estimates on the left--hand cut effects of $\pi\pi$ scatterings
in chiral perturbation theory}
\label{SecPionFF}

In above discussions we have re-expressed  the form factor and the
$S$ matrix in a way that their dependence on isolated
singularities and the $l.h.c.$ are explicitly exhibited. Since the
effects of the unitarity cut are dissolved,  the impact of the
left--hand singularities on the analytic structure of the
amplitude is singled out. These formulas can be particularly
useful  when the fine structure of the $l.h.c.$ becomes important,
which has been  ignored by most applications of the K--matrix
approach. For example, when determining the pole position of the
$\sigma$ resonance from $\pi\pi$ phase shift, since the $\sigma$
particle's mass is rather low, one has to carefully take into
account the fine structure of the $l.h.c.$, as
being emphasized recently in Ref.~\cite{AN00,ML}.
Inspired by this we in the
following carefully discuss the effects of the left--hand
singularities. In order to achieve this and to reduce the model
dependence as much as possible we need  a method to estimate the
$l.h.c.$ of $\pi\pi$ scattering amplitudes in a reliable way.
The chiral perturbation theory ($\chi$PT) affords us such a method.
The $\chi$PT treats the $T$ matrix as a perturbative
power expansion of the external momentum and works very well at
low energies. Of course, a calculation in $\chi$PT to any finite
order in the power expansion would not reconstruct the pole
structure, therefore $\chi$PT encounter problems when there is a
resonance near the physical region on the right--hand side of the
complex plane. However we argue that its predictive power in the
region near the $l.h.c.$ would suffer much less from such a
problem since the pole position is further away from the
left--hand cut. Our strategy is clear from the above discussion:
We extract from the 1--loop $\chi$PT results of the T
matrix~\cite{GM} the term relevant to the left--hand discontinuity
to estimate the left--hand integral in Eq.~(\ref{FL2}). To be
precise, the 1--loop $\chi$PT results for $\pi\pi$ scatterings are,
\bqa
T^{I=0}&=&3A(s,t,u)+A(t,u,s)+A(u,s,t)\ ,\nonumber\\
T^{I=1}&=&A(t,u,s)-A(u,t,s)\ ,\nonumber\\
T^{I=2}&=&A(t,u,s)+A(u,t,s)
\eqa
where
\bqa
A(s,t,u)&=&{s-m_\pi^2\over f_\pi^2}+B(s,t,u)+C(s,t,u)+O(E^6)\ ,
\nonumber
\\B(s,t,u)&=&{1\over 6f_\pi^4}\{3(s^2-m_\pi^4){\bar
J}(s)+\left[t(t-u)-2m_\pi^2t+4m_\pi^2u-2m_\pi^4\right]{\bar
J}(t)\nonumber \\ && +(t\leftrightarrow u)\}\ ,
 \eqa
and the function $C$ is a polynomial of $s$, $t$ and $u$ which is
irrelevant here. The function ${\bar J}(s)$ is defined as,
\bqa
{\bar J}(s)
&=& {1\over 16\pi^2}\left [\rho{\rm ln}\left({\rho-1\over
\rho+1}\right)+2\right ]\ .
\eqa
The  partial wave projection can
be carried out:
 \be
 T_J^I(s)={1\over
32\pi(s-4m_\pi^2)} \int^{0}_{4m_\pi^2-s} dt\, P_J(1+{2t\over
s-4m_\pi^2}) T^{I}(s,t,u)
\ee
from which the discontinuity of $T$
on the left can be obtained under certain conditions~\cite{MMS1},
\be
{\rm Im}_L{T^I_J(s)}={1+(-1)^{I+J}\over 32\pi(s-4m_\pi^2)}
\int_{4m_\pi^2}^{4m_\pi^2-s} dt\, P_J(1+{2t\over s-4m_\pi^2}){\rm
Im}T^{I}_t(s,t)\,\,\, ; s\le 0\ .
\ee
However, for scattering amplitudes in chiral perturbation theory,
the left--hand cut can  be directly extracted from the
analytic expressions of partial wave amplitudes.
In the $I=J=0$ channel the result is,
 \bqa
 {\rm Im}_L T_0^0(s)&=&{1\over 1536\pi^2f_\pi^4 (s-4m_\pi^2)}
 \{ 2\cdot{\rm ln}{\sqrt{4m_\pi^2-s}-\sqrt{-s}\over
 \sqrt{4m_\pi^2-s}+\sqrt{-s}}(25m_\pi^6-6m_\pi^4s)\nonumber \\
 & &+\sqrt{-s(4m_\pi^2-s)}({7\over
3}s^2-{40\over 3}m_\pi^2 s+25m_\pi^4)\} \ .
\eqa
However, this
result is still not directly applicable here. Remember that the
function $F$ is an analytic continuation of (2 times) Re$T$, which
is the real part of $T$ defined in the physical region. Since
${\rm Im}_RT$  also develops a discontinuity on the left this part
must be subtracted from  ${\rm Im}_LT$. Therefore the correct
expression for ${\rm Im}_LF$ is,
\be\label{ImLF}
{\rm Im}_LF=2({\rm Im}_LT-{\rm Re}_L{\rm Im}_RT)\ .
\ee
The 1--loop result of ${\rm Im}_RT_0^0$
is simply obtainable using the Born term amplitude and the optical
theorem,
\be
{\rm Im}_RT_0^0=\rho \left({2s-m_\pi^2\over 32\pi
f_\pi^2}\right)^2\ ,
\ee
for which we have
${\rm Re}_L{\rm Im}_RT={\rm Im}_RT$ since $\rho$ is real on the left.
It is worth pointing out that in Eq.~(\ref{ImLF}) the main contribution
to ${\rm Im}_LF$ comes from the second term on the
$r.h.s.$,~\footnote{\label{ImTR}The necessity to
have ${\rm Im}_RT$ included in Eq.~(\ref{ImLF}) can be evidenced by
Fig.~\ref{fig02}: without the contribution from ${\rm Im}_RT$ the slope of
line A or B from $\chi$PT would be significantly smaller in magnitude. See
later text for more discussions on Fig.~\ref{fig02}.}
i.e., the term
${\rm Im}_LT$ is numerically rather small comparing with ${\rm Im}_RT$
evaluated on the $l.h.c.$ (For previous discussions on the smallness
of ${\rm Im}_LT$, see for example Ref.~\cite{OO99}.).

We  also list the results in $I=J=1$ and
$I=2,J=0$ channel from chiral perturbation theory:
\bqa
{\rm Im}_LT^1_1&=&{1\over 9216 \pi^2f_\pi^4(s-4m_\pi^2)^2}
\{(36m_\pi^6-72m_\pi^4s+16m_\pi^2s^2-s^3)\nonumber
\\&&\times \sqrt{-s(4m_\pi^2-s)}+12m_\pi^4(6m_\pi^4+13m_\pi^2s-3s^2)
\nonumber \\&&\times {\rm ln}{\sqrt{4m_\pi^2-s}-\sqrt{-s}\over
 \sqrt{4m_\pi^2-s}+\sqrt{-s}}\}\ , \\
 {\rm Im}_RT_1^1&=&\rho\left(s-4m_\pi^2\over 96\pi
 f_\pi^2\right)^2\ ;\\
{\rm Im}_LT^2_0&=&{1\over 1536 \pi^2f_\pi^4(s-4m_\pi^2)}\{{1\over
6}(6m_\pi^4-32m_\pi^2s+11s^2)\sqrt{-s(4m_\pi^2-s)}\nonumber
\\&&+2(m_\pi^6+3m_\pi^4s)\times{\rm ln}{\sqrt{4m_\pi^2-s}-\sqrt{-s}\over
 \sqrt{4m_\pi^2-s}+\sqrt{-s}}\}\ ,\\
{\rm Im}_RT^2_0&=&\rho\left({s-2m_\pi^2\over 32\pi
f_\pi^2}\right)^2\ ,
\eqa
where $f_\pi=93.3$MeV and the expressions
listed above are valid up to $O(p^4)$ term in the chiral expansion.
It is worth pointing out that at $s=m_\pi^2/2$, $4m_\pi^2$ and
$2m_\pi^2$ the $lowest$ $order$ partial wave amplitudes $T_0^0$,
$T^1_1$ and $T^2_0$ vanish respectively, as a consequence of the
Adler zero condition.
\section{The determination of the $\sigma$ resonance in $\pi\pi$
scatterings}\label{res}
\subsection{A description to the fit procedure}

Having obtained the analytic expressions of ${\rm Im}_L F$
for the partial wave amplitudes it becomes possible to use
the experimental data on $\pi\pi$ scattering phase shifts
to determine the pole positions in various channels. We fit both
IJ=00, 11 and 20 channels.
In the following we  briefly  describe the procedure and the method
for the fit:

\begin{enumerate}

\item We assume the $\pi\pi$
scattering amplitude $T$ satisfies a once subtracted dispersion
relation from physical considerations. So our left--hand integral
is once subtracted at $s=m_\pi^2/2$, $s=4m_\pi^2$ and $s=2m_\pi^2$
for IJ=00, 11 and 20, respectively. The positions of the
subtraction points are chosen only for the convenience of
discussions.

\item A dispersion integral in $\chi$PT  needs
 many  subtractions and this
is artificial because of the bad high energy behavior of $\chi$PT
amplitudes. So we always truncate the once subtracted integral
 at $-\Lambda_{\chi PT}^2 $ where $\Lambda_{\chi PT}$ ranges
from, for example,
$700$MeV to $1$GeV. We argue that the influence from the
region $|\sqrt{s}|\ge \Lambda_{\chi PT}$ is negligible to the
physics we are concerning. We realize that the value $
|s|\simeq 1$GeV
is too large  for
chiral perturbation theory to be valid.
We however take this value just
for the purpose of testing to what extent our fit results depending on
the behavior of ${\rm Im}_LF$.
We have to make sure that the mass
and the width of the resonance obtained under such an
approximation have to be insensitive to the choice of the cutoff.
To compare with  perturbation results we will also extract
 ${\rm Im}_LF$ from the unitarized T matrix, i.e., the [1,1] Pad\'e
approximant, in the I=J=0 channel.

\item In the three channels IJ=00, 11 and 20, we assume one pole
in the first two cases. In the IJ=11 channel the threshold
behavior ($T\sim \rho^3$ near threshold) is considered to reduce
one parameter (the subtraction constant). In our fit each pole
contains 4 parameters. Two of them are  related to the position of
the pole. The other two are related to the couplings of the pole
to $F$. That is the coefficient $\alpha_i\equiv
i/(2\rho(z_i)S'(z_i))$  in Eq.~(\ref{FL2}). One may relate
$\alpha_i$ to $z_i$ in the narrow resonance approximation or in
models. But for the wide resonance $\sigma$ there is no simple
relation
between the two, so we take $\alpha_i$ as completely
free as an approximation.
However, we also fit in the I=J=1 channel by treating ${\rm
Re}\alpha_i$ and/or ${\rm Im}\alpha_i$ as the one obtained from
the narrow width approximation.
\end{enumerate}

\subsection{Numerical results and discussions }
\label{NRD}

The numerical results of the fit in the IJ=11 and 20  channels are
presented below:
\begin{enumerate}

\item The I=J=1 channel: The  data are taken
from table VI of Ref.~\cite{rho}. We  find that the $l.h.c.$
effect is very small in this channel. When taking $\alpha_\rho$
for free (so it is a 4 parameter fit) we obtain $M_\rho=753$MeV,
$\Gamma_\rho=142$MeV for $\Lambda_{\chi PT}=1.0$GeV. Without
$l.h.c.$ while taking $\alpha_\rho$ for free, we have instead
$M_\rho=756$MeV, $\Gamma_\rho=144$MeV  (See Fig.~\ref{fig11} for
the fit). We also tried to fix ${\rm Re}\alpha_i$ and/or ${\rm
Im}\alpha_i$ using the narrow resonance approximation. The pole
position obtained using different approaches agree with each other
within a few MeV.  The $\chi^2$ is sensitive to different
approaches, but the pole position is not.

\item The  I=2, J=0 channel:

Since there is no resonance pole in this channel
 it is a one parameter (the subtraction constant) fit here when
 $\Lambda_{\chi PT}$ is held fixed. We notice that the I=2, J=0 channel
 favors a large value of $\Lambda_{\chi PT}$.
As we see from the line B of Fig.~\ref{fig02} the phase shift is reproduced
ideally by the $l.h.c.$ integral plus one subtraction constant.
The scattering length is estimated to be $a_0^2=-0.052$ which is
about 2$\sigma$ away from the experimental value,
$a_0^2(exp)=-0.028\pm 0.012$~\footnote{All the experimental value
and the results from chiral perturbation theory
are taken from Ref.~\cite{GO99} unless quoted explicitly.} and the most recent
value from the E865 Collaboration~\cite{E865}: $a_0^2(exp)=-0.036\pm 0.009$
The case without the subtraction constant  is also
depicted in Fig.~\ref{fig02} as line A.
The latter is equivalent to imposing
the Adler zero condition for the I=2, J=0 partial--wave amplitude,
since in here the dispersion integral is subtracted at
$s=2m_\pi^2$. The scattering length $a_0^2=-0.032$ for the latter case,
which is within 1$\sigma$ error bar comparing with the experimental value.
See also footnote~\ref{ImTR} for further information.
\end{enumerate}

In the I=J=1 channel the $\rho$ resonance almost saturates the
experimental phase shift and hence it demonstrates that the
contribution from the left--hand integral must be very small, and
it is
correctly predicted by chiral perturbation theory. In contrast,
no resonance exists in the I=2, J=0 channel. It is of
course very satisfiable to see that the contribution from the
left--hand integral
 to $\delta_0^2$  is in the right direction, and
combining the contribution from the subtraction constant it
can saturate the phase shift.

From the above discussions we see that $\chi PT$
gives satisfactory results in both the IJ=11 and 20 channel.
We may obtain an impression from the above discussion that
the chiral prediction works well at least in qualitative sense, i.e.,
the order of magnitude and the sign of the left hand integrals.
In the following we turn to discuss the more interesting case of
the I=J=0 channel.

In an earlier version of the present paper,
we take the data the same as that in Ref.~\cite{ZB94}, that is the CERN-Munich
data~\cite{Gr74,Maenner74} combining with the data from $K_{e4}$
decay near  threshold~\cite{Ro77}  obtained more than 20 years ago.
The most recent experiment on $K_{e4}$ decay performed by the
E865 Collaboration~\cite{E865} is remarkable. It affords  high statistic
data on the $\pi\pi$ scattering phase shifts near threshold from which the
scattering
length parameter $a_0^0$ can be determined with much improved accuracy:
$a_0^0=0.228\pm 0.012\pm 0.003$. We therefore incorporate the new
experimental results in our fit  below.
Here we truncate the CERN-Munich data
at 900 MeV in an attempt to reduce the pollution
from the $f_0(980)$ resonance. Since the latter (or more cautiously,
its second sheet pole) is only a narrow resonance, its influence
should not be very important to the determination of the $\sigma $ resonance.
As we see from table~1 the
pole position is not very sensitive to $\Lambda_{\chi PT}$. Including
the $l.h.c.$ contribution does reduce the mass and width of the $\sigma$
particle, but the effect is not dramatic.
The $\chi^2$ are  good
for all cases, see  Fig.~\ref{fig00} for the fit.
\begin{table}[bt]
\label{tab:sigmanew1}
\centering\vspace{0.1cm}
\begin{tabular}{|c|c|c|}
\hline
&$M_\sigma$ &$\Gamma_\sigma$
\\ \hline
$\Lambda_{\chi PT}=1000$& 477&564
\\ \hline
$\Lambda_{\chi PT}=850$&503 & 587
\\ \hline
$\Lambda_{\chi PT}=700$&519 & 605
\\ \hline
Pad\'e& 519 & 579
\\ \hline
no $l.h.c.$&544 &607
\\ \hline
\end{tabular}
\caption{The pole position of the $\sigma$ resonance obtained by a
fit of the data from
Refs.~\protect\cite{Gr74,Maenner74,Ro77} and \protect\cite{E865}.
The scattering length $a_0^0$ obtained
in the global fit ranges from $0.25\sim 0.26$. The data in
the table are in units of MeV.}
\end{table}
It is very impressive to notice
that in the I=J=0 channel the contribution from the left--hand
integral to $\sin(2\delta_0^0)$ has the wrong sign
comparing with the experimental value. To clearly show this we draw
in Fig.~\ref{fig00} the contribution  from the left--hand integral
alone to $\sin(2\delta_0^0)$ in several cases, the curves are always
negative and
concave regardless of the different choice of cutoff
parameters.~\footnote{This is in qualitative agreement with the result of
Ishida et. al.~\cite{ishida} from a linear $\sigma$ model
calculation.}
The procedure presented above clearly demonstrates
the existence
of the $\sigma$ meson, in a model independent way~\footnote{
To be more careful, one may add:
provided that a chiral perturbation theory estimate on the $l.h.c.$
is qualitatively correct.}
since within our scheme the `background phase' or the cut becomes
calculable, from the principle of maximal analyticity no other
 contribution except
the $\sigma$ pole can be introduced to fit the experimental data. The
subtraction constant, which represents the contribution
from high energies, can not generate the convex curvature
that the experimental value of $\sin (2\delta_\pi)$ exhibits.

A cautious reader may worry about the convergence
problem of  the chiral expansion
in the large $\Lambda_{\chi PT}$ region, even though the qualitative
behavior of the left hand integrals obtained from perturbation theory can be
examined by varying the cutoff parameter. To overcome the difficulty we also
estimated ${\rm Im}_LF$ in a `unitarized approach', though crossing symmetry
is no longer maintained  in such an approximation.
That is we use the [1,1]
Pad\'e approximation to obtain the unitarized scattering amplitude
(in which the standard values of
the $L_i$ parameters of the Gasser--Leutwyler
Lagrangian are used) and extract
the quantity ${\rm Im}_LF$ on the left,
and insert the obtained  ${\rm Im}_LF$ into Eq.~(\ref{FL2}).
The left hand integral evaluated from the unitarized amplitude
is also plotted in Fig.~\ref{fig00} and we find that no major
conclusion is changed.
It is worth pointing out that such a
use of the Pad\'e approximation
is different from the standard usage of unitarization
(see for example Ref.~\cite{truong}--\cite{OOR00}).
Essentially we only need
the information of the unitarized amplitude on the cuts
evaluated on the left, and
the two approaches
are not technically equivalent.

However, the scattering length parameter $a_0^0$ obtained
in the global fit given in table~1
ranges from $0.25\sim 0.26$, which is about (or more than)
2$\sigma$ away from the newly
obtained experimental result (It may be worth noticing that without
the new $K_{e4}$ data the $a_0^0$ parameter obtained in our procedure
would be much larger, $a_0^0=0.35\sim 0.37$.).
To remedy this we
also include in our fit program by brute force
the experimental constraint
$a_0^0=0.228\pm 0.012\pm 0.003$.
The results are given in table~2. The fit result for $a_0^0$,
ranging from $0.230\sim 0.231$, is
stable against different treatment of the left--hand integrals.

\begin{table}[bt]
\label{tabsigma2}
\centering\vspace{0.1cm}
\begin{tabular}{|c|c|c|}
\hline
&$M_\sigma$ &$\Gamma_\sigma$
\\ \hline
$\Lambda_{\chi PT}=1000$& 454&648
\\ \hline
$\Lambda_{\chi PT}=850$&479 & 658
\\ \hline
$\Lambda_{\chi PT}=700$&500 & 662
\\ \hline
Pad\'e& 498 & 642
\\ \hline
no $l.h.c.$&525 &654
\\ \hline
\end{tabular}
\caption{The pole position of the $\sigma$ resonance with
the new experimental constraint on $a_0^0$. The value
of $a_0^0$ in the global fit ranges from $0.230\sim 0.231$.}
\end{table}

By comparing  table~1 with table~2
we find that the pole position of the $\sigma$ resonance is sensitive to
the scattering length parameter, but the influence is not as dramatic as in the case
without the new $K_{e4}$ data
from the E865 Collaboration: the pole position would be much more sensitive to the
scattering length parameter in the latter case.

In discussions above we limit ourselves to a single channel analysis,
therefore
the possible influence from higher resonances, especially
the $f_0(980)$, is omitted. We point out here that within the single channel
approximation
 the influence of
the $f_0(980)$ resonance to the fit can also be estimated (to be
exact, in a model dependent way) using the method of
Ref.~\cite{lmz}, by subtracting the phase contributed from $f_0(980)$
alone. The mass and the width of the $\sigma$ particle
change only slightly when $f_0(980)$ is taken into account,
but the qualitative role of the $l.h.c.$ is still unchanged, that
is it has only a mild influence.

As we can see from  tables~1 and 2
that
the pole position of the $\sigma$ resonance on the complex
$s$ plane, $s_\sigma$, moves towards left once the value of $a_0^0$ is
decreased~\footnote{In the previous version without the newest
$K_{e4}$ data, the fit value of Re[$s_\sigma$] is much more sensitive
to the value of $a_0^0$. Even a negative  Re[$s_\sigma$] can be obtained
when $a_0^0(\chi PT)$ is used in the fit.}.
The phenomenon that the real part of the pole position of the
$\sigma$ resonance is small has been noticed and  investigated by
Anisovich and Nikonov~\cite{AN00}. An interesting mechanism
was proposed to explain this phenomenon by suggesting
a strong singularity associated with the $l.h.c.$, which is simulated
by a series of poles on the
negative real axis fixed by the N/D  equation and the experimental
data.
Our calculation confirms that the inclusion of the $l.h.c.$ effect pushes
the $\sigma$ pole towards left but the effect is mild, and
Re[$s_\sigma$] is also found to be sensitive to the value of the
scattering length parameter.
The  difference between our treatment on the $l.h.c.$ and that
of Anisovich and Nikonov is that
the left--hand singularity of the T matrix
in our case
comes solely from the absorptive singularity, $i.e.$, the 2$\pi$
cuts in crossed channels. Due to
relativistic kinematics, the quantity $F$ contains an additional
contribution from the $s$ channel absorptive singularity,
which is also from the
$2\pi$ cut.~\footnote{\label{LHC} Related  discussions on
the left--hand cuts of $\pi\pi$ scattering amplitude
 in the recent literature may be found in
Refs.~\cite{DP97} and \cite{BP}.}
It may be illuminating to quote from Ref.~\cite{Oehme}, ``There are
$no$ such singularities which are associated with the quark--gluon
structure of hadrons, since there are no absorptive thresholds related
to this structure".

\section{Conclusion}
\label{SecConcl}
The method proposed in this paper to discuss $\pi\pi$
interactions
is based on a dispersion relation set up for the analytic
continuation of the real part
of the scattering T matrix. The
main formula Eq.~(\ref{FL2}),
though very simple, is shown to be very useful in clarifying
different contributions from poles or the left--hand cut to
the scattering phase shift. The procedure in deriving
Eq.~(\ref{FL2}) fully respects
analyticity, unitarity and crossing symmetry.

Our estimate on the left--hand cut of the scattering
T matrix obeys the
standard rule of $S$ matrix theory, that is the left--hand
singularity
comes solely from physical absorptive
singularities in the crossed channel.
In here they are the 2$\pi$ cuts from $t$ and $u$ channels. For the
quantity $F$, it also contains the part originated
from the $s$ channel
absorptive singularity, as an effect of the relativistic kinematics.

We have carefully  examined
the reliability in using chiral perturbation theory to estimate
the left--hand cut effects in various channels.  Chiral perturbation
theory encounters the problem of a bad high energy behavior,
especially in the I=J=0 channel. This
drawback was partially corrected by using a `unitarized
approach' to the perturbation series.
It is remarkable to notice
that the $l.h.c.$ contribution to $\sin (2\delta^0_0)$ is negative
and concave which clearly demonstrates the existence of the $\sigma$
resonance, according to the principle of maximal analyticity.

We have shown that the left--hand cut
effects are mild in determining the pole position of the $\sigma$
resonance, and the effect of the
scattering length in the determination of the $\sigma$ pole position
is also clarified. The estimated central value of
the mass and the width of the
$\sigma$ meson are found to be 498 MeV and 642 MeV, respectively, if
we are allowed to quote from the Pad\'e solution with the
experimental constraint on $a_0^0$. The new experimental data from the
E865 Collaboration on $K_{e4}$ decay is found to be crucial in minimizing
the uncertainty caused by $a_0^0$ in the determination of the $\sigma$ pole
position within our scheme.

The current procedure can be extended to discuss the more general
case of coupled--channel system where the $f_0(980)$ resonance has to
be taken into account~\cite{XZ}. No major conclusion on the  pole position
of the $\sigma$ resonance and the role of  the left--hand cut and the
scattering length parameter is changed.

\noindent {${\bf Acknowledgement}$:} We would like to thank Milan
Locher, Chuan-Rong Wang, Bingsong Zou
for valuable discussions. Especially we are in debt to Valeri Markushin
for his long and patient discussions,
part of sec.~2 is written by him. We also thank Qin Ang
for his assistance in performing the Pad\'e approximation.
The work of H.~Z. is supported in part by National Natural Science
Foundation of China under grant No.~19775005.
\appendix

\section{Appendix}\label{appendix}

By definition  \be S=1+2i\rho T\ee and from unitarity relation
\be{\rm Im} T=T\rho T^*\ ,\ee
We can get the analytic continuation of T on the second sheet by
using  the reflection property,
\be
T^I(s+i\epsilon)-T^{II}(s+i\epsilon)=2i\rho(s+i\epsilon)
T^I(s+i\epsilon)T^{II}(s+i\epsilon)\ ,\ee  or more concisely,
 \be
T^{II}={T^I \over 1+2i\rho T^I}={T^I \over S^I}\ . \ee
Suppose the
integration over the infinite circle is zero and using Cauchy's
theorem we get
\bqa T(s)&=&{1\over 2\pi i}\int_{C_R}{T(s')\over
s'-s}ds'+ {1\over 2\pi i}\int_{C_L}{T(s')\over s'-s}ds'+\sum_j
{\beta_j\over s-s_j}\nonumber\\ &=&{1\over
\pi}\int_{4m_\pi^2}^{\infty}{{\rm Im} T(s')\over s'-s}ds'+{1\over 2\pi
i}\int_{-\infty}^0{disc T(s')\over s'-s}ds'+\sum_j {\beta_j\over
s-s_j} \nonumber\\ &=& {1\over \pi}\int_{4m_\pi^2}^{\infty}
{\rho(s')T(s')T^*(s')\over s'-s}ds'+{1\over 2\pi
i}\int_{-\infty}^0 {disc T(s')\over s'-s}ds'+\sum_j {\beta_j\over
s-s_j}\ .
\eqa
On the $r.h.s.$ we have
\bqa
\rho(s)T^I(s)T^{I*}(s)&=&\rho(s+i\epsilon)T^I(s+i\epsilon)
T^{II}(s+i\epsilon),\nonumber\\
&=& -\rho(s-i\epsilon)T^{II}(s-i\epsilon)T^{I}(s-i\epsilon)\ ,
\eqa
therefore we
have
 \bqa \label{T} T(s)&=&{1\over 2\pi}\int_{C_R}
{\rho(s')T(s')T^{II}(s')\over s'-s}ds'+{1\over 2\pi
i}\int_{-\infty}^0 {disc T(s')\over s'-s}ds'+\sum_j {\beta_j\over
s-s_j} \nonumber\\ &=&{1\over 2\pi}\int_{C}
{\rho(s')T(s')T^{II}(s')\over s'-s}ds'-{1\over 2\pi}\int_{C_L}
{\rho(s')T(s')T^{II}(s')\over s'-s}ds' \nonumber\\ & &+{1\over
2\pi i}\int_{-\infty}^0 {disc T(s')\over s'-s}ds'+\sum_j
{\beta_j\over s-s_j} \nonumber\\ &=&{1\over 2\pi i}\int_{C}
{T(s')({S^I-1 \over 2S^I})\over s'-s}ds'+{1\over 2\pi
i}\int_{-\infty}^0 {disc( T(s'){S^I+1\over 2S^I})\over
s'-s}ds'+\sum_j {\beta_j\over s-s_j} \cr &=&\Phi+{1\over 2\pi
i}\int_{-\infty}^0 {disc( T(s'){S^I+1\over 2S^I})\over
s'-s}ds'+\sum_j {\beta_j\over s-s_j}
 \eqa
where
 \bqa \Phi &\equiv
& {1\over{2\pi i}} \int_{C}{T(s')({S^I-1 \over 2S^I})\over
s'-s}ds' \nonumber\\&=& T({S^I-1\over
2S^I})+\sum_i{1/2i\rho(s_i)\over{2S^I (s_i)}'(s_i -s)}-\sum_j
{\beta_j\over 2(s-s_j)}\ .
 \eqa
 In the derivation of
Eq.~(\ref{T}), by combining the second and the third term on the
right hand side of the second equation of (\ref{T}) and using \be
\rho T^{II}={S^I-1\over 2iS^I}\ee we  get the third equation of
(\ref{T}). And by substituting $\Phi$ into (\ref{T}) we  obtain
\be
 T(s)={S\over S+1}\left[\sum_i {i\over 2
\rho(s_i){S(s_i)}'(s-s_i)}+\sum_j{\beta_j\over s-s_j}+{1\over 2\pi
i}\int_{-\infty}^0 {disc( T(s'){S+1\over S})\over s'-s}ds'\right]
\ .
\ee

\begin{newpage}
\begin{figure}[hbtp]
\begin{center}
\vspace*{-10mm} \mbox{\epsfysize=70mm\epsffile{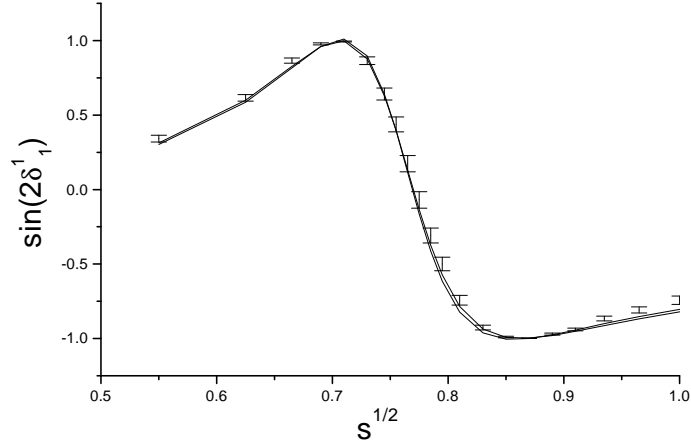}}
\vspace*{0mm} \caption{\label{fig11} The fit in the I=J=1 channel
with and without $l.h.c.$ effects.}
\end{center}
\end{figure}
\begin{figure}[hbtp]
\begin{center}
\vspace*{-10mm} \mbox{\epsfysize=70mm\epsffile{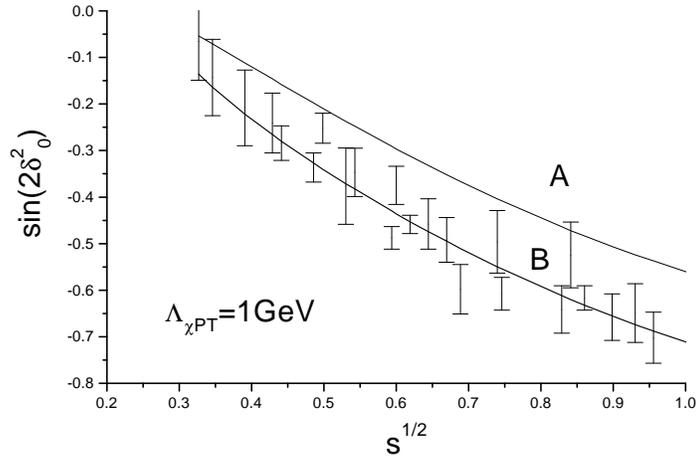}}
\vspace*{0mm} \caption{\label{fig02}
Contributions to the
scattering phase from the left--hand integral (once subtraction at
$s=2m_\pi^2$ is used) in the  $I=2,J=0$ channel. Line A
corresponds to setting $T^2_0(s=2m_\pi^2)=0$ and be parameter
free. and line B corresponds to a free subtraction constant fixed
by minimizing $\chi^2$. The data are from Ref.~\protect\cite{Gr74}.}
\end{center}
\end{figure}
\begin{figure}[hbtp]
\begin{center}
\vspace*{-10mm} \mbox{\epsfysize=90mm\epsffile{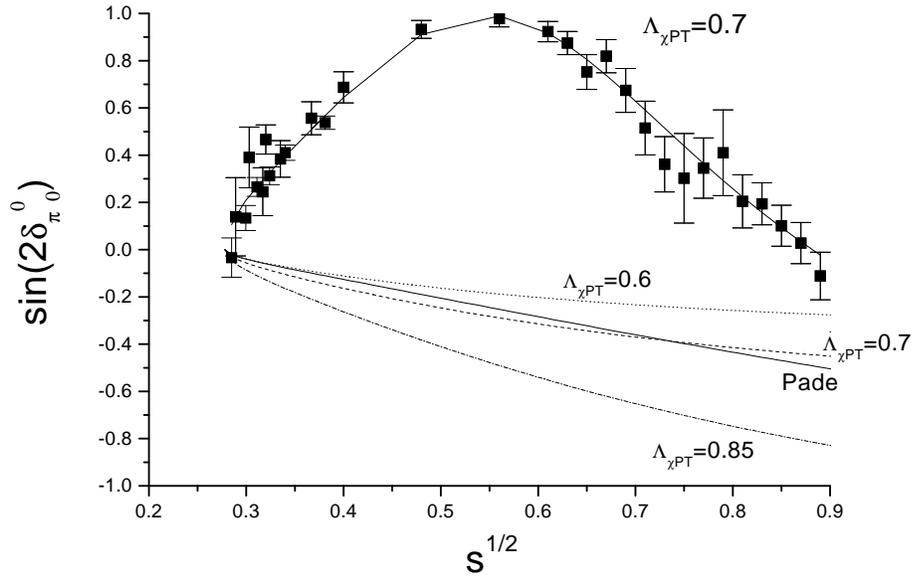}}
\vspace*{0mm} \caption{\label{fig00}A typical fit of 5 parameters
(4 resonance + 1 subtraction constant) in the I=J=0 channel, with
$\Lambda_{\chi PT}$=0.7GeV and the experimental constraint on $a_0^0$.
Different estimates on the left--hand integral are also plotted: The
dotted line corresponds to $\Lambda_{\chi PT}=0.6 GeV$, the dashed line corresponds to
$\Lambda_{\chi PT}=0.7 GeV$, the dot--dashed line corresponds to $\Lambda_{\chi PT}=0.85 GeV$
and the solid line corresponds to the Pad\'e solution.}
\end{center}
\end{figure}
\end{newpage}
\end{document}